# Modeling EEG data distribution with a Wasserstein Generative Adversarial Network to predict RSVP Events

Sharaj Panwar, Paul Rad, Tzyy-Ping Jung, Yufei Huang*

*Abstract*—Electroencephalography (EEG) data are difficult to obtain due to complex experimental setups and reduced comfort with prolonged wearing. This poses challenges to train powerful deep learning model with the limited EEG data. Being able to generate EEG data computationally could address this limitation. We propose a novel Wasserstein Generative Adversarial Network with gradient penalty (WGAN-GP) to synthesize EEG data. This network addresses several modeling challenges of simulating time-series EEG data including frequency artifacts and training instability. We further extended this network to a class-conditioned variant that also includes a classification branch to perform event-related classification. We trained the proposed networks to generate one and 64-channel data resembling EEG signals routinely seen in a rapid serial visual presentation (RSVP) experiment and demonstrated the validity of the generated samples. We also tested intra-subject cross-session classification performance for classifying the RSVP target events and showed that class-conditioned WGAN-GP can achieve improved event-classification performance over EEGNet.

## I. INTRODUCTION

ELECTROENCEPHALOGRAPHY (EEG) is an attractive neuroimaging tool for measuring brain activities due to its portability, noninvasiveness and its ability to capture spatiotemporal dynamics of human brains. However, obtaining high-quality EEG data could be labor-intensive and costly. The scarcity of high-quality EEG data poses significant challenges in the era of deep learning (DL) to train high-performing deep models to predict cognitive events and understand associated brain dynamics and mechanisms. It is thus of great interest in developing cost-effective approaches to augment the limited EEG samples so that the superb ability of DL in learning data representation can be fully exploited for EEG-based cognitive event classification.

Generative Adversarial Networks (GANs) are a class of deep generative models that can learn to generate samples from the data distribution [1]. The basic, or vanilla, GAN consists of a generator and a discriminator, both of which are deep neural networks. The generator tries to fool the discriminator by producing realistic fake samples, and the discriminator tries to distinguish between the real and fake samples [2]. The generator and discriminator are optimized together according to a two-player minimax game. GAN has been recently applied in a variety of applications from generating photorealistic images of human faces, to image style transferring, and to data augmentation [3]–[7]. In addition to serving as deep generative models, the GAN framework has also been extended to solve many other learning problems including domain adaption [8] and semi-supervised learning [9]–[15]. Such a GAN model architecture can also be useful for class-conditioned learning [16], [17], where the class labels are used in generating samples from a specific class and a classifier can also be added to the discriminator to predict the labels of an unlabeled sample. Intuitively, this approach augments the training samples with GAN generated samples. Once trained, no sample generation is needed and the classifier exploits the superior ability of GAN in learning data representations to achieve improved performance. GAN based algorithms for class-conditioned learnings are an active area of research, with many new improvements being proposed [13], [14], [17]–[21].

Motivated by these advancements of GAN, this study aims to explore GAN architectures for generating synthetic EEG data and design class-conditioned architectures for improving the performance of EEG-based cognitive event classification without the need to collect additional samples. The convolutional neural network (CNN) and recurrent neural network (RNN) have been extensively used for creating EEG based discriminative models. They can also be used as building blocks within a generative framework such as GAN to generate EEG samples. The proposed GAN framework indeed uses CNNs in both generator and discriminator. A recent model called EEG-GAN for generating a single-

The research was supported by a grant from Intel Corporation and also sponsored by the Army Research Laboratory under Cooperative Agreement Number W911NF-10-2-0022. The views and conclusions contained in this document are those of the authors and should not be interpreted as representing the official policies, either expressed or implies, of the Army Research Laboratory or the U.S. Government. The U.S. Government is authorized to reproduce and distribute reprints for the Government purposes notwithstanding any copyright notation herein.

*Corresponding author: Yufei Huang (email: yufei.huang@utsa.edu)

Yufei Huang is with Department of Electrical and Computer Engineering, The University of Texas at San Antonio, San Antonio, Texas, United States.

Sharaj Panwar is also with Department of Electrical and Computer Engineering, The University of Texas at San Antonio, San Antonio, Texas, United States.

Paul Rad is with Department of Information System, The University of Texas at San Antonio, San Antonio, Texas, United States.

Tzyy-Ping Jung is with Swartz Center for Computational Neuroscience, University of California, San Diego, San Diego, CA, United States.

channel EEG data from a motor task was proposed in [22]. This study evaluated several upsampling methods in the generator architecture for their ability to address the frequency artifacts in the generated EEG samples. The paper suggests using interpolation as a preferred method for the upsampling task over deconvolution due to its better ability to control frequency artifacts. No single interpolation method stands out to be the best because the quantitative evaluation and visual inspection led to contradictive conclusions. The study also cautioned extra care in interpreting the results from quantitative evaluation methods such as the inception score, because they could end up favoring a model that generates a distribution away from the real distribution [20][21]. As not a single metric could provide sufficient information about the quality of the generated samples, visual inspection is indispensable and quite often is still the most reliable practice. Two other recent work [25][26] addressed the problem of generating multi-channel EEG samples but both in the context of data augmentation for classification of cognitive tasks. Although a recurrent neural network was proposed for the generator in [26] to model temporal dynamics of EEG signals and a conditional Wasserstein GAN was introduced in [25] to improve the stability of GAN training, no attention is given to the influence of different upsampling approaches on the artifacts in the generated samples and careful quality assessment of the generated samples is lacking. While both studies showed that augmenting existing training samples with GAN-generated samples improved classification performance for emotion recognition and motor imagery tasks, sample generation and classification were treated as separate tasks. As a result, GAN was not fully exploited for classification because the generated samples were not optimized to improve the classification performance.

To address these issues in the existing work, this study proposed a novel WGAN model with a gradient penalty specifically designed for generating high-quality multi-channel EEG data. This study carefully examined the impact of different upsampling approaches on signal frequency and amplitude. Study results confirmed the fact as reported in EEG-GAN that deconvolution resulted in signal artifacts. However, the results also revealed that interpolation considerably degraded the signal amplitude even though it is better than deconvolution in preserving the signal frequency. To overcome these problems, this study proposed and evaluated a two-step upsampling approach with a combination of bi-cubical interpolation, followed by deconvolution with bi-linear weight initialization. Based on this novel WGAN model, we also proposed a class-conditioned WGAN model that performs event classification in addition to sample generation. Because the sample generation is trained to optimize the classification, it is designed to improve the classification performance using training samples alone.

The remaining of the paper is organized as follows. Section II introduces the background on Wasserstein Generative Adversarial Networks with gradient penalty (WGAN-GP) and class-condition WGAN-GP (CC-WGAN-GP). We discuss the proposed upsampling scheme and provide the details of the proposed architectures for WGAN-GP and CC-WGAN-GP. Section III presents the experimental results on evaluating the performance of the proposed WGAN in generating samples from single- and multi-channel EEG data. Cross-session prediction results by CC-WGAN-GP are also discussed. Section IV summarizes the main results and concludes this study.

## II. METHODS

### A. Background on Wasserstein Generative Adversarial Networks

*1) Vanilla WGAN-GP:* The GAN is designed to learn to generate samples from the distribution of the training data. It consists of two deep networks that try to outplay each other [2]. Given the training, or "real", data distribution $\mathbb{P}_r$ and the generated, or "fake", data distribution $\mathbb{P}_f$, the discriminator network $D$ with parameters $\emptyset_D$ is trained to distinguish between the real $x_r$ and fake $x_f$ data. The second deep network, the generator $G$ with parameter $\emptyset_G$ takes a latent noise sample $z$ from the distribution $\mathbb{P}_z$ as input and is trained to generate fake samples $x_f$ to fool the discriminator $D$. This results in a minimax game, in which the $G$ is forced by the $D$ to produce better samples. The vanilla GANs suffer heavily from training instability and mode collapsing and thus are restricted to generating low resolution samples [27]. Much advancement in training stability and the sample quality has been made in recent years [26, 27]. The Wasserstein GAN with a gradient penalty (WGAN-GP) loss function has been shown to effectively improve the training stability and convergence [27]. The WGAN loss computes the Wasserstein distance, also called the earth moving distance between the real and fake data distributions [28]:

$$\overline{W}(\mathbb{P}_r, \mathbb{P}_f) = E_{x_r \sim \mathbb{P}_r}[D_{\emptyset_D}(x_r)] - E_{x_f \sim \mathbb{P}_f}[D_{\emptyset_D}(x_f)]. \quad (1)$$

The discriminator is trained to minimize the Wasserstein distance for a fixed generator or $\emptyset_G^*$

$$L_D(\emptyset_D, \emptyset_G^*) = \overline{W}(\mathbb{P}_r, \mathbb{P}_f) \quad (2)$$

whereas the generator is trained to maximize the following loss with a fixed $\emptyset_D^*$

$$L_G(\emptyset_G, \emptyset_D^*) = E_{x_f \sim \mathbb{P}_f}[D_{\emptyset_D^*}(x_f)] = E_{x_f \sim \mathbb{P}_f}[D_{\emptyset_D}(G_{\emptyset_G^*}(z))] \quad (3)$$

The WGAN requires $D$ to be K-Lipschitz [28], which can be achieved by clipping the weights of the discriminator to be inside an interval $[-c, c]$. To better enforce the Lipschitz continuity on the discriminator, a solution is achieved by adding the following gradient penalty to the WGAN loss (1)[27]

$$\widetilde{W}(\mathbb{P}_r, \mathbb{P}_f) = \overline{W}(\mathbb{P}_r, \mathbb{P}_f) + \lambda E_{\hat{x} \sim \mathbb{P}_{\hat{x}}}[(\|\nabla_{\hat{x}} D(\hat{x})\|_2 - 1)^2] \quad (4)$$

where $\lambda$ is a hyper-parameter controlling the trade-off between the WGAN loss and gradient penalty, $\hat{x}$ denotes the samples lying on a straight line between $\mathbb{P}_r$ and $\mathbb{P}_f$.

*2) **Class-Conditioned WGAN-GP (CC-WGAN-GP) Classifier:*** The WGAN framework can be extended for supervised learning from the training data with class labels $y_r$ by replacing the discriminator with the discriminator/classifier as shown in Figure 1[15][18]. Specifically, the discriminator/classifier share the lower layers $S$ with parameters $\emptyset_S$ but have a unique discriminator branch $D$ with parameters $\emptyset_D$ and a classifier branch $C$ with parameters $\emptyset_C$ at the output. Adding a classifier branch does not entirely change the basic WGAN framework and the discriminator still plays the same role in the GAN training. However, the generator $G$ becomes the class conditioned WGAN, which generates samples $x_f$ conditioned on the class labels $y$ in addition to the random sample.

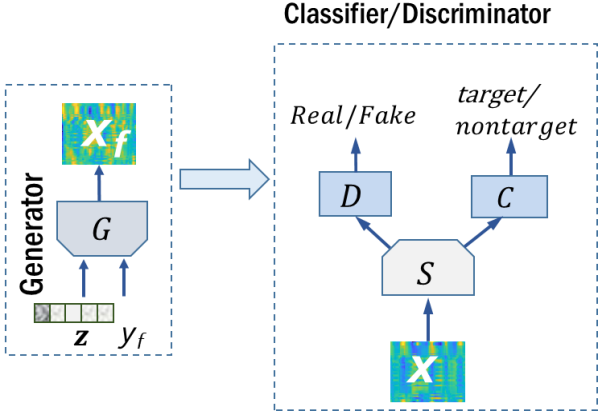

**Figure 1**. *Class-Conditioned WGAN-GP*. Generator $G$ takes an input $z$ and the class label $y_f$ to produce a sample $x_f$. The classifier/discriminator takes a sample $x$ being either a fake sample $x_f$ or a real sample $x_r$ as the input and feed it to the shared architecture $S$ before discriminating the status (real/fake) of the input samples for the discriminator $D$ and predicting the class labels $y$ of the input epoch with the classifier $C$.

The use of class labels for improving GAN training has proven to be very effective for training stability and controlling the generator output [16]. A popular such model is the auxiliary GAN (ACGAN) [17], which uses the auxiliary class labels in the generator and perform the classification task. We adopt a similar idea in our proposed CC-WGAN-GP. However, AC-GAN used the cross-entropy loss for both the discriminator and classifier, whereas we used the Wasserstein distance for the discriminator loss and the cross-entropy for the classifier. Specifically, the discriminator branch and the shared layers of CC-WGAN-GP are trained to minimize the WGAN-GP loss for a fixed $\emptyset_G^*$ as

$$L_D(\emptyset_G^*, \emptyset_D, \emptyset_S) = \widetilde{W}(\mathbb{P}_r, \mathbb{P}_f). \quad (6)$$

The class labels $y_f$ for the fake samples are also used as target labels for the classifier branch to optimize it on the generated data $x_f$. The classification branch now maximizes the likelihood of classes $y_r$ and $y_f$ for the real and fake samples respectively for a fixed $\emptyset_G^*$ as

$$L_C(\emptyset_G^*, \emptyset_C, \emptyset_S) = E[\log p(y_r|x_r)] + E[\log p(y_f|x_f)] \quad (7)$$

where $x_f = G_{\emptyset_G^*}(z, y_f)$. Equivalently, the discriminator/classifier branch is trained to minimize the following combined loss

$$L_{D/C}(\emptyset_G^*, \emptyset_D, \emptyset_C, \emptyset_S) = L_D(\emptyset_G^*, \emptyset_D, \emptyset_S) - L_C(\emptyset_G^*, \emptyset_C, \emptyset_S). \quad (8)$$

The generator is then trained to maximize the following modified loss for fixed $\emptyset_D^*, \emptyset_C^*,$ and $\emptyset_S^*$

$$L_G(\emptyset_G, \emptyset_D^*, \emptyset_C^*, \emptyset_S^*) = E_{x_f \sim \mathbb{P}_f}\left[D_{\emptyset_D^*}\left(G_{\emptyset_G}(z, y_f)\right)\right] + L_C(\emptyset_G, \emptyset_C^*, \emptyset_S^*) \quad (9)$$

Note from (6) and (7) that this training of the classifier resembles augmenting classification training with GAN generated samples as proposed in [25][26]. However, instead of training the GAN separately from the classifier, the proposed CC-WGAN-GP trains the classifier as part of the GAN, thus fully exploiting GAN's ability in learning data representations for improved classification performance.

### B. The proposed generator upsampling layer for EEG time-series data

This section discusses our proposed generator design for accurately generating EEG signals. The GAN generator takes

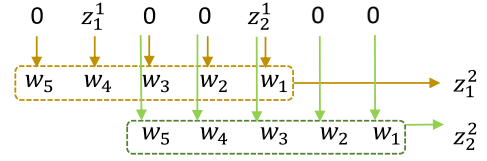

**Figure 2**. Illustration of the deconvolution process of upsampling the input $z^1$ to upsampled output $z^2$ by a factor of 3

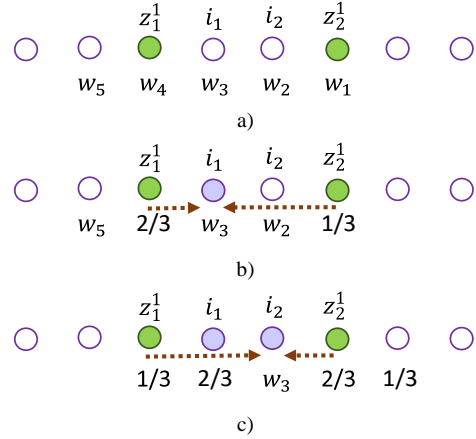

**Figure 3** a) Bilinear weights $\{w_1, w_2, w_3, w_4, w_5\}$ Initialization of a deconvolution kernel for an upsampling factor 3, the corresponding. Input data points $(z_1^1, z_2^1)$ and the data point for the linear interpolation $i_1$ and $i_2$. b) Initializing weights $w_4$ and $w_1$ to perform interpolate at pixel '$i_1$'. c) Initializing weights $w_5$ and $w_2$ to perform interpolate at pixel '$i_2$' a vector of random noise as input and generates samples by gradually upsampling the input dimension. The upsampling method plays an important role in determining the quality of generated samples, which in turn impacts overall GAN training. The upsampling task is also a key feature of the GAN architecture for the generation of time-series EEG data because it controls the artifacts in the generator and directly affects the stability of the discriminator and classifier training. Deconvolution, also known as transposed convolution or fractionally strided convolution, has been effectively used for increasing the resolution in image-related tasks and can also be used here for upsampling in the GAN generator [29]. In the deconvolution operation, based on an upsampling factor, zeros are inserted between consecutive input values and a trainable kernel is applied to perform a convolution operation as shown in Fig. 2. The number of zeros to be inserted is calculated by subtracting 1 from the upsampling factor. However, when the kernel stride of the deconvolution is smaller than the kernel size, there is an uneven overlap in the output regions [30]. The uneven overlap positions cause the checkerboard pattern in the resulting image [31]. Ideally, the weights should be learned for unevenly overlapping positions so that the checkerboard patterns at the output can be mitigated. However, it is a balancing act to learn the weights for both evenly and unevenly overlapping positions. This problem can be partially addressed by choosing a kernel size that is divisible by the deconvolution stride but evenly overlapping regions can also cause the learned kernels to produce similar artifacts. A significant presence of such patterns in time series signals produces huge frequency artifacts, which can force the discriminator to learn a trivial solution to reject the generated samples based on these artifacts and ruin the entire training process. One natural approach to avoid these artifacts is to unfold the deconvolution process by using an upsampling layer through linear interpolation, cubical interpolation, or nearest-neighbor interpolation followed by a convolutional kernel [32].

As will be seen in Section III, interpolation methods such as the nearest neighbor and bi-cubical interpolations tend to produce significantly reduced signal amplitudes as compared to real signals, although the generated signals can match the shape of the real EEG signals. The classifier of such a CC-WGAN model also yields both poor classification performance and longer converge times. Conversely, the deconvolution matches the amplitude but produces large artifacts in the generated signals.

To address these issues, we propose a deconvolution layer with linear weights initialization. The idea of initializing the deconvolutional kernel is to encourage a linear interpolation. The appropriate kernel size of the deconvolution kernel should be: $kernel\ size = (2 \times stride - stride\%2)$, where % computes the remainder. The linear weights initialization for one-dimensional upsampling assuming an upsampling factor of 3 and a deconvolutional kernel with weights $w_1, w_2, w_3, w_4,$ and $w_5$ is depicted in Fig. 3. The weight calculation for interpolation of $i_1$ (Fig. 2) is performed by sliding the center weight $w_3$ on $i_1$. The linear weights $w_1$ and $w_4$ can now be calculated based on the contribution of $z_1^1$ and $z_2^1$ to interpolate $i_1$, where $z_1^1$ and $z_2^1$ are the elements of input to the upsampling layer. Hence, the weight $w_4$ turns out to be 2/3 and $w_1$ to be 1/3. Similarly, by sliding the center weight $w_3$ at $i_2$, we can calculate the contributions of $z_1^1$ and $z_2^1$ for interpolating $i_2$. Finally, we can assign a weight value 1 to the center weight $w_3$. This way we can still preserve the advantages of the linear interpolation by avoiding artifacts and have a trainable kernel to help learn the amplitude of real samples and at the same time increase the classification performance.

The edges of generated samples, especially in the case of multi-channel EEG data could have lop-sided amplitude, where the generated signals in individual channels can be seen as having very high or very low amplitude at the edges [33]. Also, sometimes the generated samples are DC-shifted versions of real EEG samples. The phenomenon is likely caused by the upsampling process, especially the interpolation of EEG signal values at the edges, where the contributing signal time points are less in numbers as compared to that of a central time point. To fix this, we propose to first generate a higher resolution sample say $72 \times 72$ instead of the original $64 \times 64$ and then have a clipping layer to crop it to the dimension of the real sample. The generated samples are then normalized to zero to remove the DC shift.

*C. The proposed WGAN architectures*

This section details the architectures of the proposed WGAN models. For all the models, the generator includes a two-step upsampling by a factor of 2 with combinations of fully connected layers and convolutional layers. The first upsampling step uses a bi-cubical interpolation followed by a convolutional layer. The second step uses a deconvolution with bi-linear weights initialization. This upsampling arrangement gives the best performance in reducing artifacts and improving WGAN training and classification. The deconvolutional kernel in both cases is kept divisible by the stride in order to avoid checkerboard patterns as discussed in section II (B) and [31]. Also, for an upsampling factor of 2, the stride is kept as 2 and thus the deconvolution kernel size is calculated as $2 \times 2 - 2\%2 = 4$, or $(1 \times 4)$ kernels for one-channel EEG data and $(4 \times 4)$ for 64-channel data. We next discuss the detailed architectures for generating epochs of single-channel and 64-channel EEG signals and predicting events by the CC-WGAN classifier.

*1) WGAN-GP model for one-channel EEG signal*

Table 1 shows the detailed WGAN-GP model architecture for generating an epoch of a single-channel EEG signal. The generator takes input as a sample from a 120-dimensional i.i.d standard normal distribution followed by a fully connected layer of 1024 neurons and the Leaky Relu activation function. Another fully connected layer with 2048 neurons is added and batch normalization is performed at the output of this layer followed by the Leaky Relu. The output of the Leaky Relu is reshaped to $(128 \times 1 \times 16)$. A Bi-cubical interpolation upsampling with batch normalization and Leaky Relu is included to increase the resolution in the time dimension to $(128 \times 1 \times 32)$. The resultant output is fed to a 2D convolutional block with 64 kernels of size $(1 \times 3)$ with batch normalization and activation for the second stage of upsampling, which uses a deconvolution layer of $(1 \times 4)$ kernel with bilinear weight initialization. Finally, a convolutional layer with a single kernel of size $(1 \times 3)$ is applied at the end of the generator to generate a 2D EEG sample. since we want to generate a 2D sample consistent with real signals. The single kernel is used to keep the depth of the generated sample as 1. The discriminator takes an input dimension of $(1 \times 64)$ and has a common CNN architecture. Gaussian white noise with mean 0 and a standard deviation 0.05 is added at the beginning of discriminator architecture to help improve training stability and model learning as proposed in [24], [34]. They argue that the main source of instability stems from the fact that the real and the generated distributions have disjoint supports or lie on low-dimensional manifolds. In the case of an optimal discriminator, this will result in zero gradients that then stop the training of the generator. By adding Gaussian noise to the discriminator, we can avoid the zero gradients and help sustain the GAN training. We next add three blocks of 2D convolution with Leaky Relu and flatten the output to feed to a fully connected block with 1024 neurons. A layer with the linear activation is added in the end to produce a single neuron.

**Table 1**: The proposed architecture for one-channel EEG data

| Generator | | Discriminator | |
|---|---|---|---|
| Layer | Output Shape | Layer | Output Shape |
| Input layer | 120 | Input layer | 1, 64, 1 |
| Fully connected | 1024 | Gaussian noise | 1, 64, 1 |
| LeakyReLU | 1024 | Conv. 2D | 1, 64, 64 |
| Fully connected | 2048 | LeakyReLU | 1, 64, 64 |
| Batch Norm. | 2048 | Conv. 2D | 1, 32, 128 |
| LeakyReLU | 2048 | LeakyReLU | 1, 32, 128 |
| Reshape | 1, 16, 128 | Conv. 2D | 1, 16, 128 |
| Upsampling | 1, 32, 128 | LeakyReLU | 1, 16, 128 |
| Batch Norm. | 1, 32, 128 | Fully connected | 2048 |
| LeakyReLU | 1, 32, 128 | Fully connected | 1024 |
| Conv. 2D | 1, 32, 64 | LeakyReLU | 1024 |
| Batch Norm. | 1, 32, 64 | Fully connected | 1 |
| LeakyReLU | 1, 32, 64 | Output layer | 1 |
| Deconv. with Bilinear weights | 1, 64, 128 | **Generator Parameters:** Total: 2,285,761 | |
| Batch Norm. | 1, 64, 128 | Trainable: 4,736 | |
| LeakyReLU | 1, 64, 128 | **Discriminator Parameters:** | |
| Conv. 2D | 1, 64, 1 | Total: 2,173,441 | |
| Output layer | 1, 64, 1 | Trainable: 2,173,441 | |

**Table 2:** The proposed architecture for 64-channel EEG data

| Generator | | Discriminator | |
|---|---|---|---|
| Layer | Output Shape | Layer | Output Shape |
| Input layer | 120 | Input layer | 64, 64, 1 |
| Fully connect. | 1024 | Gaussian noise | 64, 64, 1 |
| LeakyReLU | 1024 | Conv. 2D | 64, 64, 64 |
| Fully connect. | 41472 | LeakyReLU | 64, 64, 64 |
| Batch Norm. | 41472 | Conv. 2D | 32, 32, 128 |
| LeakyReLU | 41472 | LeakyReLU | 32, 32, 128 |
| Reshape | 18, 18, 128 | Conv. 2D | 16, 16, 128 |
| Upsampling | 36, 36, 128 | LeakyReLU | 16, 16, 128 |
| Batch Norm. | 36, 36, 128 | Fully connect. | 32768 |
| LeakyReLU | 36, 36, 128 | Fully connect. | 1024 |
| Conv. 2D | 36, 36, 64 | LeakyReLU | 1024 |
| Batch Norm. | 36, 36, 64 | Fully connect. | 1 |
| LeakyReLU | 36, 36, 64 | Output layer | 1 |
| Deconv. with Linear weight | 72, 72, 128 | **Generator Parameters:** Total: 34,049,601 | |
| Clip layer | 64, 64, 128 | Trainable: 33,983,425 | |
| Batch Norm. | 64, 64, 128 | **Discriminator Parameters:** | |
| LeakyReLU | 64, 64, 128 | Total: 33,777,536 | |
| Conv. 2D | 64, 64, 1 | Trainable: 33,777,536 | |
| Output layer | 64, 64, 1 | | |

*2) 64-channel WGAN-GP:*

The main difference between the 64-channel and 1-channel architectures is that for the 64-channel (Table 2) the operations such as upsampling, convolution, and activations are performed in both EEG channel and time dimensions. The dimension of the convolutional kernel is kept as $(3 \times 3)$ and the dimension of the deconvolutional kernel with bi-linear weights is $(4 \times 4)$. To generate 64-channel signals, the number of neurons of the second fully connected layer (Table 2) is increased from 2048 to 41,472 and after applying a batch normalization and Leaky Relu, the output is reshaped to $18 \times 18$ as shown in Table 2. As discussed in section II, we generate a higher dimension $(72 \times 72)$ signal and then crop it to $(64 \times 64)$ with the customized clipping layer. The discriminator architecture is very similar to the one used in the one-channel case except that we use $(3 \times 3)$ convolution kernels and apply all operations such as batch normalization in both the time and channel dimensions.

*3) 64-Channel CC-WGAN-GP Classifier:*

The CC-WGAN-GP classifier has a similar model architecture as the 64-channel WGAN-GP. Since the input to the generator is class-conditioned, the randomly generated labels are embedded in the input noise vector. An embedded layer is added at the input of the generator for this purpose. Since the goal is to train a classifier simultaneously with the WGAN training, the normalization is done during the training process within the generator itself. A customized normalization layer is also added at the output of the generator. The normalized output is sent to the network $S$ shared by the discriminator and classifier (Fig. 1), which has the same network structure as the 64-channel vanilla WGAN-GP except that it has two output branches for discrimination and classification. The discriminator has the same architecture as before while the classifier branch has a fully connected layer with two neurons that predict the likelihood of a sample belonging to a target or a non-target class.

*D. Evaluation methods for the generated samples*

The evaluation of GAN performance and sample quality is an open problem. There are multiple quantitative measures including inception score, Frechet inception score, Euclidian distance, and sliced Wasserstein distance, each often limited in their scope [35]. The EEG-GAN has also cautioned the use of these quantitative evaluation approaches to access the quality of generated EEG as they often give contradicting evaluation and can lead to a model that generates a distribution far from the real one [22], [23], [36], [37].

The inception score based evaluation matrices are widely used to access the quality of image generation where an efficient inception classifier trained on the ImageNet dataset is necessary to determine the entropy of the conditional label distribution of generated samples [24]. However, for RSVP data and most of the EEG-based classification, there is a limitation with obtaining a highly accurate classifier, which would cause the inception scores inconclusive. Also, the inception score fails to give useful information about the generator and is susceptible to noise. In this case, a good quality assessment along with visual inspection is still the most effective approach for evaluating the generated EEG signals. We discuss next the approaches adopted in this study.

*1) Visual Inspection.* Visual inspection of the generated samples has proven to be one of the best ways to assess the quality of the samples in case of images. For the RSVP EEG data, one can inspect if ERP (Event-related Potential) or P300 is present in the generated target samples. This can be used to inspect whether or not the generated samples resembling the features in the real data. Fig. 4 shows an averaged P300 in target responses from a sample subject.

*2) Log-Likelihood score from Gaussian mixture models (GMMs).* The complexity of the experiment and human brain activities causes the RSVP data to have multiple modes under both target and non-target events. A key aspect of assessment is to inspect if the WGAN models capture the characteristics of the EEG signals. To this end, we propose to use the GMM to assess the modes of single-channel EEG data. Given an EEG sample $x$, we assume that it follows a GMM with $K$ components, i.e.

$$p(x) = \sum_{k=1}^{K} w_k \mathcal{N}(x|\mu_k, \Sigma_k) \qquad (10)$$

where $\mathcal{N}(\cdot \mid \cdot, \cdot)$ is the Gaussian distribution and $\mu_k$ and $\Sigma_k$ are the mean and covariance matrix of component $k$, and $w_k \in (0,1)$ is the weight of component $k$ with the constraint $\sum_{i=1}^{K} w_i = 1$. A Bayesian Information Criteria (BIC) is used for selecting the best number of mixture components $K$. To quantitatively evaluate the quality of the generated single-channel samples, we calculate the log-likelihood $\log p(x)$ for each of the generated samples before averaging them as the final log-likelihood score. The closer the score of an upsampling approach is to that of the real samples, the better the approach.

III. EXPERIMENTS AND RESULTS

The experiments for investigating the performance of the proposed WGAN-GP models are divided into four parts. We started our experiments with training a WGAN-GP model to generate a sinusoidal signal whose frequency and amplitude are known. We evaluated different upsampling methods in matching the frequency and amplitude of the sinusoid. Then, Then, we trained a WGAN-GP for the single-channel data and examined the ability of different upsampling schemes in matching the modes of the EEG signals. Next, we extended the investigation to generating 64-channel target and non-target samples. The insights gained from these experiments are used to create a class-conditioned WGAN-GP classifier. The classification performance is investigated for each subject and compared with EEGNet.[38] During the WGAN training, we kept the training ratio of the discriminator and generator to 1: 5. This arrangement gives the maximum performance regarding the amplitude and matches the morphology of the EEG signal. The area under the ROC curve (AUC) is used to measure classification performance.

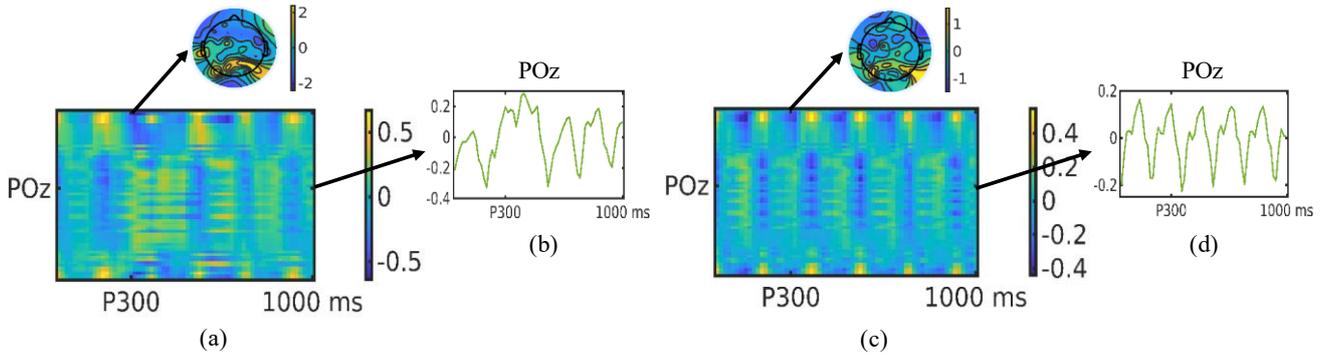

**Figure 4.** Averaged heatmaps/reponses of target/nontarget samples from session 1 of subject 1. (a) The heatmap of the target sample with the topograph at 300ms. (b) Plot of the target signal at POz channel. (c) The heatmap of the nontarget sample with the topograph at 300ms. (d) Plot of the nontarget signal at POz channel.

*A. The RSVP Dataset*

1) **The RSVP experiment.** This study used the dataset collected from the BCIT X2 rapid serial visual presentation (RSVP) experiment [32][33], where the test subjects were asked to identify rare target images from a continuous burst of image clips presented at a rate of 5 Hz. The images ($512 \times 662\ pixels$) are of 5 different objects (doors, chairs, etc.) and in each session, one of the five objects was treated as target and other objects were non-targets. Each of the 10 subjects performed 5 sessions ($\sim 1\ h\ per\ session$). EEG signals were measured by 256-channel BioSemi EEG systems.

2) **The RSVP dataset.** The preprocessing was performed using PREP [41], [42], which included band-pass filtering from 0.1 to 55 Hz, robust signal referencing, identifying and interpolating the bad channels (channels with a low recording SNR), and baseline removal using EEGLab [43]. All these preprocessed datasets were down-sampled to 64Hz and a subset of 64 channels associated with the visual cortex region were selected. The data collected from Subjects 2 and 9 did not show clear ERPs for target images and hence were excluded from this study. One-second EEG epochs after each image presentation onset were extracted and z-score normalized by the epoch mean and standard deviation. We chose the samples from session 1 of subject 1 (S1R1) for investigating the task of EEG generation and use samples from all subjects for the classification task. The heatmaps of the averaged target and non-target epochs are shown in Fig. 4(a)&(c). ERPs are visible in the target heatmap around 300ms. For the nontarget heatmap, a 5Hz pattern can be clearly seen, which is the result of the repetitive image presentation at 5Hz.

As an intermediate step of the investigation, we also examined the training of a WGAN model to generate samples from a single EEG channel. To this end, we used the sequential forward selection criteria for selecting an EEG channel with the maximum discriminate power [44]. A support vector machine (SVM) with a polynomial kernel was trained for each channel for S1R1and the POz channel was selected (Area under the ROC curve (AUC)=54.59%) as a result. The averaged signals for the target and non-target samples of S1R1 at POz are shown in Fig. 4b&d and the P300 can be clearly seen in the target epochs.

*B. Generating noisy sinusoid signals*

Assessing the pros and cons of the generator architecture directly on multi-channel EEG signals consisting of multiple unknown frequency components with varying amplitude and phase is highly challenging. Instead, to gain a fundamental understanding of how the different generator upsampling schemes could impact the generated time-series data, we first focused on generating a simple sinusoid signal with known frequency and amplitude. A toy dataset containing 5000 samples of noisy sinusoid was created, where each sample has 64 values sampled from a 1s sinusoid with a frequency of 5Hz and an amplitude of 1 corrupted with additive white Gaussian noise with variance 1. Experiments were designed to understand the effect of various upsampling methods on capturing the frequency and amplitude of this simple sinusoid. The upsampling approaches including nearest-neighbor interpolation (NN) and bi-cubic interpolation (BC) proposed in EEG-GAN [22], deconvolution (DC), and deconvolution with bilinear weights initialization (DCBL) and their combinations were examined for the two-step upsampling process. The upsampling combinations including deconvolution followed by another deconvolution (DC-DC), EEG-GAN proposed bicubic interpolation followed by bicubic interpolation (EEG-GAN-BCBC) and nearest neighbor followed by the nearest neighbor (EEG-GAN-NNNN), bicubic interpolation followed by a deconvolution with bilinear weights initialization (BC-DCBL) and deconvolution with bilinear weights initialization followed by another deconvolution with bilinear weights initialization (DCBL-DCBL) were examined. Fig. 5 shows the results. As can be seen, the deconvolution combination (DC-DC) created considerable very low-amplitude artifacts mainly due to the "checkerboard effect" of the deconvolution [31]. On the other hand, the interpolation-based upsampling methods (EEG-GAN-BCBC; EEG-GAN-NNNN) could match the frequency of the sinusoid but failed to generate the correct amplitude. In contrast, the proposed combination of bi-cubical interpolation followed by deconvolution with bi-linear weights initialization (BC-DCBL) outperformed all other approaches and correctly modeled both the frequency and amplitude of the sinusoid.

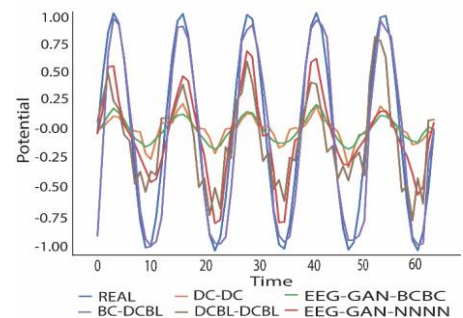

**Figure 5.** Averaged real sinusoid and generated sinusoids by various upsampling combination.

*C. Generating one-channel EEG data:*

The experiments performed in the previous section on sinusoid provide insights into the ability of various upsampling methods in matching sinusoid frequency and amplitude, giving a strong motivation to use BC-DCBL. However, it is still important to investigate how these methods perform in generating more complicated and realistic EEG data with multiple modes in their distributions. Here, we focused on three upsampling approaches from the previous section (BC-DCBL, DCBL-BC, BC-BC), which were shown to generate comparatively better sinusoidal waveforms. These approaches are examined for generating EEG signals from a single channel. For this purpose, we trained two WGAN-GP models for target and non-target signals separately using samples from the POz channel of S1R1.

To quantitatively evaluate of generated samples, we first fitted the real target and non-target EEG samples with GMM and then calculated the Log-likelihood distances between the real EEG and GMM samples ($Real{\sim}GMM$) as well as the generated samples and GMM samples ($Gen{\sim}GMM$) by different generator architectures i.e. upsampling methods.

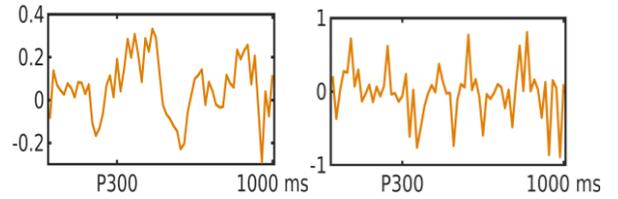

**Figure 6.** (a) Plot of averaged generated target POz samples. (b) Plot of average generated non-target POz samples.

**Table 3:** GMM log-likelihood scores for real and generated target and non-target samples of POZ by various methods

| | GMM Comp. | Real | BC-DCBL | DCBL-BC | BC-BC |
|---|---|---|---|---|---|
| Target | 4 | -66.07 | -54.48 | -44.04 | -43.51 |
| Non-Target | 3 | -30.44 | -43.75 | -56.32 | -49.32 |

BC-DCBL has a likelihood distance closet to that of ($Real{\sim}GMM$) for both generated target samples and the difference of the log-likelihood between $Real{\sim}GMM$ and $Gen{\sim}GMM$ is significantly less for BC-DCBL than those from other approaches for both target (*p* value 2.3e-10; t-test) and non-target data (*p* value 1.4e-10; t-test) as shown in Table 3.

Finally, we inspected the averaged generated target and non-target samples from the BC-DCBL architecture, which has shown so far to provide the best performance. The averaged generated target and non-target POz samples are shown in Fig 6(a)&(b). The similarity between these results and the real averaged target and non-target samples in Fig. 4 (b)&(d) is evident. The ERP can be seen clearly in the generated target sample and the 5Hz presentation-associated pattern is also visible in the non-target epoch. Taken together, the BC-DCBL scheme has shown to provide the best performance in matching the frequency and amplitude of the sinusoid and learning the modes, ERP, and signal patterns of the one-channel EEG signals.

*D. Generating 64-channel EEG data*

Based on the results from the previous sections, we trained the WGAN-GP models to generate 64-channel EEG data

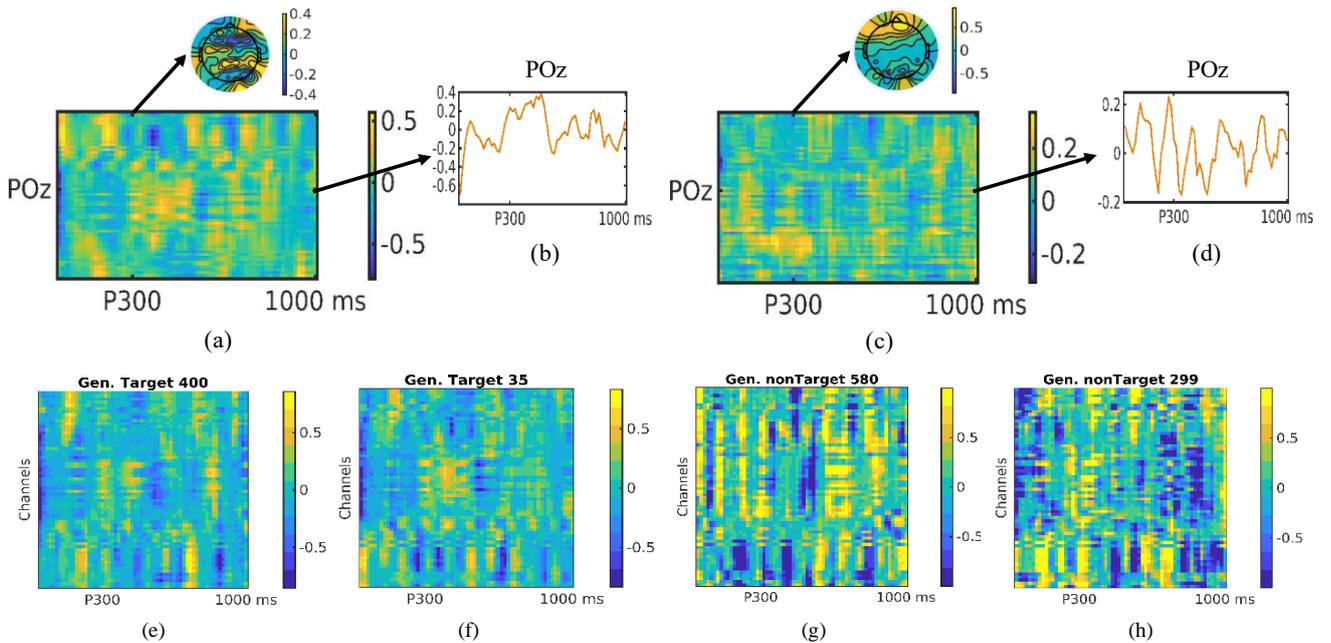

**Figure 7** (a) Heatmap of the generated target sample with the topograph at 300ms. (b) Plot of the generated target signal at the POz channel. (c) Heatmap of the generated nontarget sample with the topograph at 300ms. (d) Plot of the generated nontarget signal at the POz channel. (e)-(f) randomly selected examples of generated target samples. (g)-(h) randomly selected examples of generated nontarget samples.

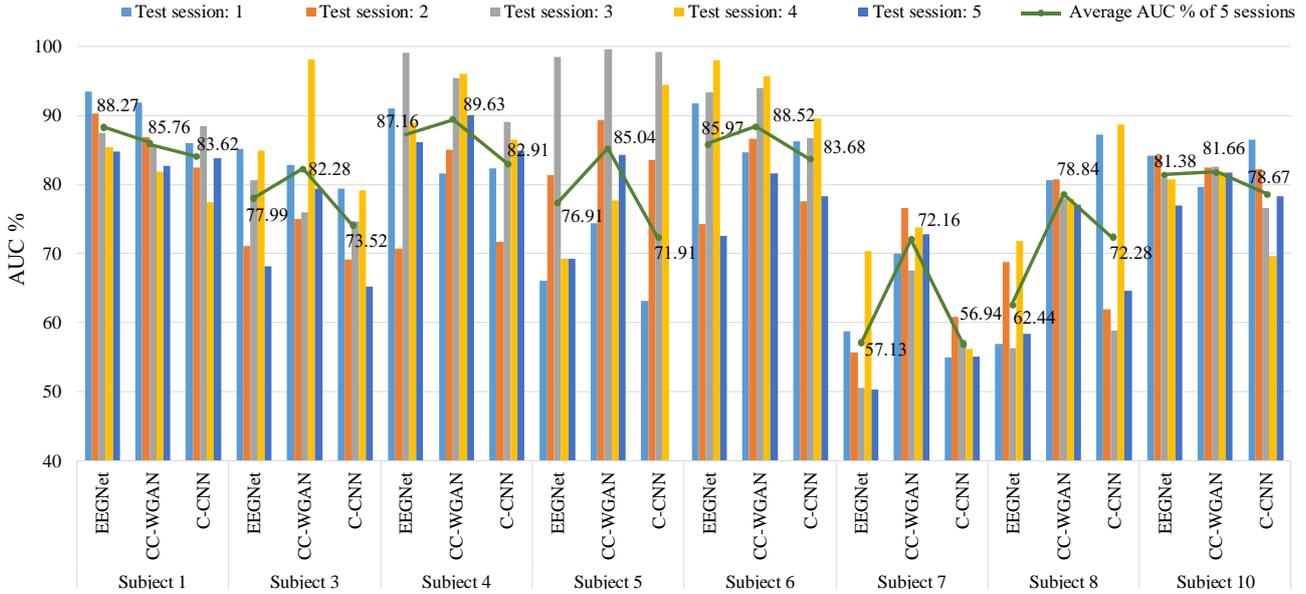

**Figure 8** Cross-session classification performance for 8 subjects.

using the BC-DCBL upsampling scheme. The proposed generator includes the BC-DCBL upsampling layers with 3x3 convolutional kernels. Two separate WGAN-GP models were trained for the target and nontarget samples. To evaluate the quality of the generated samples, visual inspection was performed by comparing the heatmaps of the averaged generated samples with those of the real samples. We also compare the generated target and non-target samples from the POz channel with the real samples. A topograph at P300 is plotted for both the generated target and non-target samples to show event-related potentials. The heatmaps of target and non-target samples are depicted in Fig. 7(a)&(c), respectively. We can clearly see the P300 of the ERPs in the target epoch and even detailed ERP patterns in the real target heatmap in Fig. 4(a). The 5Hz oscillations associated with the repetitive image presentation are also visible in both target and non-target epochs, although these oscillation patterns are not as smooth and sharp as in the real nontarget epoch in Fig. 4(c). The topography of the generated samples for the generated target sample at 300ms also confirmed that ERPs were predominately from the occipital region. In contrast, the prefrontal region is activated in the non-target sample around 300ms. We also plotted the averaged generated target and nontarget signals at POz channel as depicted in Fig. 7(b)&(d) and they again captured the ERP and image presentation patterns in the real signals. Randomly selected examples of generated target and nontarget samples are provided in Fig. 7(e,f)&7(g,h), respectively. We further investigated the effect of gradient penalty term on sample generation and generated samples using vanilla GAN without GP. The average GMM log-likelihood scores for each of the 64 channels were calculated and the distance from the scores of real samples averaged over the 64 channels is reported in Table 4. We can clearly see that the samples generated from vanilla GAN-GP are much closer to thoes of the real samples. The large difference of GAN without GP suggests that GP is necessary to capture multiple modes in the data distribution. Taken together, these results clearly demonstrate the ability of the trained WGANs-GP in capturing important signal features in the generated EEG signals

**Table 4:** Average differences of GMM log-likelihood scores bwtween for real and generated target and non-target samples from 64-channel GAN models

|  | GAN-GP | GAN | CC-WGAN-GP |
| --- | --- | --- | --- |
| Target | 5.39 | 10.44 | 11.68 |
| Non-Target | 10.57 | 14.54 | 14.58 |

### E. Classification of RSVP events with class-conditioned WGAN-GP

Once we validated the performance of WGAN-GP in generating EEG signals, we extended the architecture to train the CC-WGAN-GP for RSVP target classification. The classification experiments in this section were performed on all 5 recording sessions of the 8 subjects. To investigate the performance, we focused on the intra-subject cross-session classification, where 5-fold cross-validation was performed for each subject and for each fold, data from 4 sessions were used for training and those from the 5th session were used for testing. Because we were interested in the classification performance, to assess the training convergence, the checkpoints were created where the AUC performance of the CC-WGAN-GP classifier was evaluated after each training epoch and the model with the maximum classification performance was retained in the end for prediction.

The results were compared with EEGNet and shown in Fig. 8. Except for subject 1, the CC-WGAN-GP outperformed EEGNet in terms of averaged AUC for each of the subjects. The improvement ranges from 0.28% (subject 10) to 16.4% (subject 8), with an average improvement of 5.83% across subjects. At the session level, out of 40 cross-session tests, CC-WGAN-GP obtained an averaged performance of 82.98±7.68, compared to EEGNet's 77.16±13.59. This result suggests that CC-WGAN-GP has not only a significantly higher per-session AUC than EEGNet (*p*-value 0.01, one-side t-test) but also a more robust performance due to a small AUC standard deviation. Furthermore, as expected, CC-WGAN-GP has greater AUC improvements in cases where EEGNet has a lower performance. Interestingly, the AUC improvement has a significant negative correlation with the EEGNet AUC (Pearson correlation: -0.84; Fig. 9).

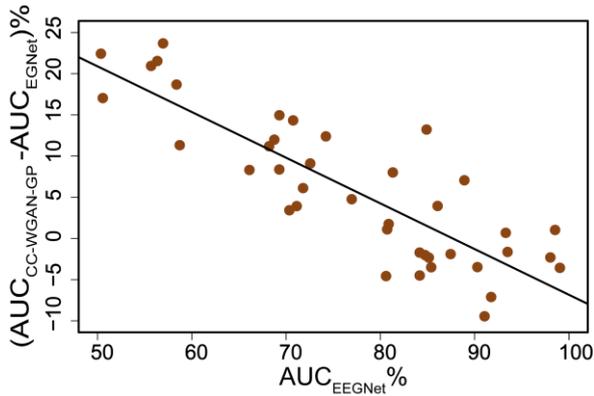

**Figure 9.** Scatter plot of AUC improvement ($AUC_{CC-WGAN-GP} - AUC_{EEGNet}$)% vs. % AUC of EEGNet for 40 test sessions. There is a significant negative correlation (Pearson correlation: -0.84).

The classifier of CC-WGAN-GP is deeper and more complex than EEGNet. Then, we wonder if increase performance of CC-WGAN-GP is due to this increase in the classification architecture. To this end, we trained a baseline classification network (C-CNN) using the CC-WGAN-GP classifier architecture. The AUC performance of the C-CNN classifier is shown in Fig 9. We see that the average performance of C-CNN is 75.44±14.71 and is worse than EEGNet (77.16±13.59). This result shows that the improved performance of CC-WGAN-GP is mainly due to the GAN-based data augmentation by the generator. The fact that C-CNN has a larger model size than EEGNet suggests that C-CNN overfitted the data. However, this increased classifier model complexity is necessary for CC-WGAN-GP to capture the additional information in GAN-based data augmentation.

To further understand the role of the generator on the classification performance of CC-WGAN-GP, we investigated the generated samples by the CC-WGAN-GP generator. We calculated the GMM log-likelihood scores on the samples generated for subject 1 session 1. The average log-likelihood score distance to the real samples is reported in Table 4. Interestingly, CC-WGAN-GP samples have the biggest differences with the real samples when compared with GAN-GP and GAN, suggesting that they are unlikely from the high-density regions of the data distribution. It has been shown in [14] that the generated samples that can reduce the classification loss would come from the low-density region of the data distribution. Our result confirms this finding. Augmenting these generated samples, which are under-represented in training data, improves the representation of the real data distribution and hence improves the classification performances, especially on subjects that the EEGNet did not perform well. This is a compelling result that substantiates the exact motivation for us to take advantage of the superb ability of GAN in learning the data representation to help improve classification performance.

Taking these results together, we conclude that the proposed CC-WGAN-GP can improve the classification performance of the supervised DL models and result in more robust predictions.

## IV. DISCUSSION AND CONCLUSION

A systematic study was performed to investigate the GAN model for generating time-series EEG responses and for performing semi-supervised EEG-based event classification. Wasserstein-GAN with a gradient penalty was investigated for generating single-channel EEG data and 64-channel EEG data. The models were created using a wide range of experiments and by carefully inspecting the GAN training over its ability to model complex brain activities measured by EEG. The upsampling step is crucial for generator architecture. We observed that the interpolation alone resulted in much-reduced amplitude in the generated samples, slowed down the training convergence, and eventually caused the training to diverge, thus incapable of generating quality samples. To address this problem, we proposed a two-step upsampling approach with a combination of bicubic interpolation followed by deconvolution with bilinear weight initialization. This proposed approach was demonstrated to be able to generate samples with the desired amplitude and mitigate the frequency artifacts, thus resulting in improved GAN training.

Another phenomenon observed in the generated samples was the lop-sided amplitude, i.e., irregular amplitudes on the edges of generated samples. The phenomenon was addressed by first generating a higher dimension EEG sample in the upsampling layer and then cropping it in later layers using a customized clipping layer to output the generated samples with the dimension of the real sample. This trick helped the training and improved the quality of the generated samples. The additional DC shift was further removed by normalizing the generated samples over their mean.

Evaluating the quality of the generated EEG samples with multiple modes is still an open research topic. As discussed earlier, the existing quantitative evaluation matrices often produce inconclusive results because of ill-satisfied conditions. We proposed a log-likelihood score based on mixture Gaussian models to access the quality of generated samples. This proposed approach aims at capturing the modes of the sample distributions and was shown to produce quantitative results consistent with a visual inspection.

Another key contribution of the paper is the proposed Class-Conditioned Wasserstein Generative Adversarial Network Classifier with the gradient penalty (CC-WGAN-GP). The proposed model includes a classifier in WGAN-GP to perform task classification. In addition, the class label is also included in the generator, which helps alleviate mode collapse [16], important design consideration for generating EEG signals with intricate patterns. Two recent work in [25] and [26] proposed to augment the training EEG samples with GAN generated samples and showed that this GAN-based augmentation approach could improve the classification performance. However, because GAN was trained independent of the classification task, the generated samples cannot be guaranteed always to help improve the classification. Also, determining the optimal number of augmented samples associated with the best classification performance is nontrivial and could complicate the process of classification training. In contrast, one key advantage of our proposed CC-WGAN-GP over existing GAN-based data augmentation approaches is that sample generation and classification are trained together with a unified loss so that the generated samples are optimized for improved classification performance. Test results on 40 intra-subject, cross-session prediction of RSVP target events demonstrated robust performance improvement over the state-of-the-art EEGNet. The improvement was especially pronounced for cases when EEGNet's performance was low.

We believe that this work contributes to advancing the current GAN research for generating EEG data and improving EEG-based event classification. This work has the potential for further expansion into related applications such as EEG super-resolution, semi-supervised learning, transfer learning, and domain adoption.

REFERENCES

[1] Y. Lu, Y.-W. Tai, and C.-K. Tang, "Attribute-Guided Face Generation Using Conditional CycleGAN," May 2017.
[2] I. Goodfellow *et al.*, "Generative Adversarial Nets." pp. 2672–2680, 2014.
[3] M. Frid-Adar, I. Diamant, E. Klang, M. Amitai, J. Goldberger, and H. Greenspan, "GAN-based Synthetic Medical Image Augmentation for increased CNN Performance in Liver Lesion Classification," Mar. 2018.
[4] F. H. K. dos S. Tanaka and C. Aranha, "Data Augmentation Using GANs," Apr. 2019.
[5] H. Shi, L. Wang, G. Ding, F. Yang, and X. Li, "Data Augmentation with Improved Generative Adversarial Networks," in *2018 24th International Conference on Pattern Recognition (ICPR)*, 2018, pp. 73–78.
[6] J.-Y. Zhu, T. Park, P. Isola, A. A. Efros, and B. A. Research, "Unpaired Image-to-Image Translation using Cycle-Consistent Adversarial Networks Monet Photos," 2018.
[7] T. Karras, T. Aila, S. Laine, and J. Lehtinen, "Progressive Growing of GANs for Improved Quality, Stability, and Variation," Oct. 2017.
[8] E. Tzeng, J. Hoffman, K. Saenko, and T. Darrell, "Adversarial Discriminative Domain Adaptation."
[9] J. T. Springenberg, "Unsupervised and Semi-supervised Learning with Categorical Generative Adversarial Networks," Nov. 2015.
[10] T. Miyato, A. M. Dai, and I. Goodfellow, "Adversarial Training Methods for Semi-Supervised Text Classification," May 2016.
[11] A. Kumar, P. Sattigeri, and P. T. Fletcher, "Semi-supervised Learning with GANs: Manifold Invariance with Improved Inference."
[12] Y. Tu, Y. Lin, J. Wang, and J.-U. Kim, "Semi-Supervised Learning with Generative Adversarial Networks on Digital Signal Modulation Classification," *CMC*, vol. 55, no. 2, pp. 243–254, 2018.
[13] A. Odena, "Semi-Supervised Learning with Generative Adversarial Networks," Jun. 2016.
[14] Z. Dai, Z. Yang, F. Yang, W. W. Cohen, and R. Salakhutdinov, "Good Semi-supervised Learning that Requires a Bad GAN," May 2017.
[15] S. Panwar, P. Rad, J. Quarles, E. Golob, and Y. Huang, "A semi-supervised wasserstein generative adversarial network for classifying driving fatigue from eeg signals," in *Conference Proceedings - IEEE International Conference on Systems, Man and Cybernetics*, 2019, vol. 2019–October, pp. 3943–3948.
[16] M. Mirza and S. Osindero, "Conditional Generative Adversarial Nets," Nov. 2014.
[17] A. Odena, C. Olah, and J. Shlens, "Conditional Image Synthesis With Auxiliary Classifier GANs," Oct. 2016.
[18] C. Wang, C. Xu, X. Yao, and D. Tao, "Evolutionary Generative Adversarial Networks," Mar. 2018.
[19] Q. Xie, Z. Dai, E. Hovy, M.-T. Luong, and Q. V. Le, "Unsupervised Data Augmentation for Consistency Training," Apr. 2019.
[20] D. Berthelot, N. Carlini, I. Goodfellow, N. Papernot, A. Oliver, and C. Raffel, "MixMatch: A Holistic Approach to Semi-Supervised Learning," May 2019.
[21] S. Panwar, P. Rad, J. Quarles, and Y. Huang, "Generating EEG signals of an RSVP experiment by a class conditioned wasserstein generative adversarial network," in *Conference Proceedings - IEEE International Conference on Systems, Man and Cybernetics*, 2019, vol. 2019–October, pp. 1304–1310.
[22] K. G. Hartmann, R. T. Schirrmeister, and T. Ball, "EEG-GAN: Generative adversarial networks for electroencephalograhic (EEG) brain signals," p. 7, Jun. 2018.
[23] M. Heusel, H. Ramsauer, T. Unterthiner, B. Nessler, and S. Hochreiter, "GANs Trained by a Two Time-Scale Update Rule Converge to a Local Nash Equilibrium." pp. 6626–6637, 2017.
[24] T. Salimans *et al.*, "Improved Techniques for Training GANs." pp. 2234–2242, 2016.
[25] Y. Luo and B.-L. Lu, "EEG Data Augmentation for Emotion Recognition Using a Conditional Wasserstein GAN," in *2018 40th Annual International Conference of the IEEE Engineering in Medicine and Biology Society (EMBC)*, 2018, vol. 2018, pp. 2535–2538.
[26] S. M. Abdelfattah, G. M. Abdelrahman, and M. Wang, "Augmenting The Size of EEG datasets Using Generative Adversarial Networks," in *2018 International Joint Conference on Neural Networks (IJCNN)*, 2018, pp. 1–6.
[27] I. Gulrajani, F. Ahmed, M. Arjovsky, V. Dumoulin, and A. C. Courville, *Improved Training of Wasserstein GANs*. 2017, pp. 5767–5777.
[28] M. Arjovsky, S. Chintala, and L. Bottou, "Wasserstein GAN," Jan. 2017.
[29] A. Radford, L. Metz, and S. Chintala, "Unsupervised Representation Learning with Deep Convolutional Generative Adversarial Networks," Nov. 2015.
[30] J. G.-C. P. for S. CS231N and U. 2014, "Conditional generative adversarial nets for convolutional face generation," *pdfs.semanticscholar.org*, vol. 2, 2014.
[31] A. Odena, V. Dumoulin, and C. Olah, "Deconvolution and Checkerboard Artifacts," *Distill*, vol. 1, no. 10, p. e3, Oct. 2016.
[32] C. Dong, C. C. Loy, K. He, and X. Tang, "Image Super-Resolution Using Deep Convolutional Networks," *IEEE Trans. Pattern Anal. Mach. Intell.*, vol. 38, no. 2, pp. 295–307, Feb. 2016.
[33] F. Fahimi, Z. Zhang, W. B. Goh, K. K. Ang, and C. Guan, "Towards EEG Generation Using GANs for BCI Applications."
[34] I. Sutskever, R. Jozefowicz, K. Gregor, D. Rezende, T. Lillicrap, and O. Vinyals, "Towards Principled Unsupervised Learning," Nov. 2015.
[35] A. Borji, "Pros and cons of GAN evaluation measures," *Comput. Vis. Image Underst.*, vol. 179, pp. 41–65, Feb. 2019.
[36] G. Peyré and M. Cuturi, "Computational Optimal Transport," *Found. Trends® Mach. Learn.*, vol. 11, no. 5–6, pp. 355–206, 2019.
[37] J. Rabin, G. Peyré, J. Delon, and M. Bernot, "Wasserstein Barycenter and Its Application to Texture Mixing," Springer, Berlin, Heidelberg, 2012, pp. 435–446.
[38] V. J. Lawhern, A. J. Solon, N. R. Waytowich, S. M. Gordon, C. P. Hung, and B. J. Lance, "EEGNet: A Compact Convolutional Neural Network for EEG-based Brain-Computer Interfaces," 2018.
[39] J. Touryan, G. Apker, B. J. Lance, S. E. Kerick, A. J. Ries, and K. McDowell, "Estimating endogenous changes in task performance from EEG," *Front. Neurosci.*, vol. 8, Jun. 2014.
[40] J. Touryan, B. J. Lance, S. E. Kerick, A. J. Ries, and K. McDowell, "Common EEG features for behavioral estimation in disparate, real-world tasks," *Biol. Psychol.*, vol. 114, pp. 93–107, Feb. 2016.
[41] N. Bigdely-Shamlo, T. Mullen, C. Kothe, K.-M. Su, and K. A. Robbins, "The PREP pipeline: standardized preprocessing for large-scale EEG analysis," *Front. Neuroinform.*, vol. 9, p. 16, Jun. 2015.
[42] Z. Mao, W. X. Yao, and Y. Huang, "Design of deep convolutional networks for prediction of image rapid serial visual presentation events," in *2017 39th Annual International Conference of the IEEE Engineering in Medicine and Biology Society (EMBC)*, 2017, pp. 2035–2038.
[43] A. Delorme and S. Makeig, "EEGLAB: an open source toolbox for analysis of single-trial EEG dynamics including independent component analysis," *J. Neurosci. Methods*, vol. 134, no. 1, pp. 9–21, Mar. 2004.
[44] Hanchuan Peng, Fuhui Long, and C. Ding, "Feature selection based on mutual information criteria of max-dependency, max-relevance, and min-redundancy," *IEEE Trans. Pattern Anal. Mach. Intell.*, vol. 27, no. 8, pp. 1226–1238, Aug. 2005.